\documentclass[%
reprint,amsmath,amssymb,aps,prl,superscriptaddress,longbibliography,
floatfix,
nobalancelastpage
]{revtex4-2}

\usepackage{graphicx}% Include figure files
\usepackage{dcolumn}% Align table columns on decimal point
\usepackage{bm}% bold math
\usepackage{epstopdf}
\usepackage{float}
\usepackage{braket}
\usepackage{dsfont}
\usepackage{amsmath}
\usepackage{siunitx} % typesets numbers with units very nicely
\usepackage[italicdiff]{physics}
\usepackage[T1]{fontenc}
\usepackage[dvipsnames]{xcolor}
\usepackage{color,soul}
\usepackage{lipsum}
\usepackage[colorlinks=true, allcolors=blue]{hyperref}
\usepackage{float}
\usepackage{placeins}

\newcommand{\pwisein}{\left\{ \begin{array}{ll}}
\newcommand{\pwiseout}{\end{array}\right.}

\newcommand{\dez}{\Delta E_Z}

\begin{document}

\title{Dynamically corrected gates in silicon singlet-triplet spin qubits}

\author{Habitamu Y.Walelign}
\thanks{These authors contributed equally.}
\author{Xinxin Cai}
\thanks{These authors contributed equally.}
\affiliation {Department of Physics and Astronomy, University of Rochester, Rochester,NY 14627}

\author{Bikun Li}
\affiliation{Department of Physics, Virginia Tech, Blacksburg, Virginia 24061}
\affiliation{Pritzker School of Molecular Engineering, The University of Chicago, Chicago, IL 60637, USA}

\author{Edwin Barnes}
\affiliation{Department of Physics, Virginia Tech, Blacksburg, Virginia 24061}
\affiliation{Virginia Tech Center for Quantum Information Science and Engineering, Blacksburg, VA 24061, USA}

\author{John M. Nichol}
\email{john.nichol@rochester.edu}
\affiliation {Department of Physics and Astronomy, University of Rochester, Rochester,NY 14627}

\date{\today}
\begin{abstract}
Fault-tolerant quantum computation requires low physical-qubit gate errors. Many approaches exist to reduce gate errors, including both hardware- and control-optimization strategies. Dynamically corrected gates are designed to cancel specific errors and offer the potential for high-fidelity gates, but they have yet to be implemented in singlet-triplet spin qubits in semiconductor quantum dots, due in part to the stringent control constraints in these systems. In this work, we experimentally implement dynamically corrected gates designed to mitigate hyperfine noise in a singlet-triplet qubit realized in a Si/SiGe double quantum dot. The corrected gates reduce infidelities by about a factor of three, resulting in gate fidelities above $0.99$ for both identity and Hadamard gates. The gate performances depend sensitively on pulse distortions, and their specific performance reveals an unexpected distortion in our experimental setup. 
\end{abstract}

\maketitle

\section{Introduction}
Quantum computers are predicted to solve some problems exponentially faster than classical computers~\cite{shor1994algorithms,ekert1996quantum}.  One important challenge in realizing this potential is that single- and multi-qubit gate fidelities must be high enough to enable use of error-correcting codes.  For example, the surface code has been shown to tolerate gate infidelities in the range of $0.6\% - 1\%$~\cite{fowler2009high,raussendorf2007fault}. Moreover, the resource requirements of error correction schemes increase with error rates. Thus, implementing gates with high fidelities is critically important for quantum computers.  

There are many different approaches to eliminate gate errors. Hardware-level approaches can involve altering the qubit design or fabrication to suppress noise and decoherence. In superconducting systems, for example, adding a large capacitance in parallel to a Josephson junction mitigates charge noise~\cite{koch2007charge}, and in semiconductor spin qubits isotopic purification reduces hyperfine noise~\cite{burkard2023semiconductor}. Because such hardware modifications are often challenging, control approaches are an attractive alternative to improving coherence and fidelities. For example, parameter regimes that feature reduced sensitivity to noise, or ``sweet spots,'' offer one route to increased coherence and fidelity~\cite{reed2016reduced, martins2016noise}. Other strategies, like dynamical decoupling, involve additional pulses to refocus qubit states.

While effective, the control methods discussed above have limitations. Sweet spots require operating qubits in specific parameter regimes, and dynamical decoupling approaches usually only implement identity gates. Another control approach involves dynamical error correction through the careful design of control fields to achieve a target operation while at the same time canceling errors to known sources of noise. Such dynamically corrected gates (DCGs) offer the prospect of significantly increased gate fidelities at the cost of precise pulse control~\cite{khodjasteh2009dynamical,barnes2015robust}. There are many highly-developed theoretical formalisms for the construction of such gates~\cite{khodjasteh2009dynamically,biercuk2009optimized,khodjasteh2010arbitrarily,wang2012composite,kestner2013noise,zeng2018general,huang2019high}. However experimental demonstrations of DCGs are limited. Dynamically corrected gates have been implemented in nitrogen-vacancy centers in diamond~\cite{rong2014implementation} and in single-electron spin qubits in silicon~\cite{yang2019silicon}, and related work has demonstrated the suppression of infidelity due to $1/f$ electrical noise in spin qubits~\cite{huang2017robust}. Although theoretical proposals for implementing DCGs in singlet-triplet qubits have existed for some time~\cite{wang2012composite,kestner2013noise}, DCGs have yet to be demonstrated in these systems. This is due in part to constraints on exchange pulses, which must be real and nonnegative for quantum dots like ours with strong electronic confinement and in small magnetic fields~\cite{li2010exchange,wagner1992spin}. This constraint on exchange couplings, which essentially results from the Lieb-Mattis theorem, limits the applicability of standard nuclear magnetic resonance sequences~\cite{wang2012composite} or numerical optimal control methods such as GRAPE~\cite{khaneja2005optimal}.

In this work, we design, implement, and assess dynamically corrected identity and  Hadamard gates on a silicon singlet-triplet qubit to correct for hyperfine noise. We use the recently-developed space curve quantum control (SCQC) formalism~\cite{barnes2022dynamically} to design our control pulses.  Crucially, the control pulses feature a small number of parameters, which we can systematically vary to calibrate the gates. Through process tomography, we find that the DCGs reduce infidelities by about a factor of three, compared with standard uncorrected gates. As expected, the performance of the gates depends sensitively on pulse distortions and charge noise in our system, as confirmed by numerical simulations. Based on these numerical simulations, we hypothesize the presence of a partially broken gate in our system, which leads to an unexpected pulse distortion. 

In total, our results establish the potential for high-fidelity operations with corrected gates generated via the SCQC formalism. While we demonstrate suppression of hyperfine errors, these gates can in principle correct for various forms of noise while respecting qubit control constraints~\cite{nelson2023designing}.  Corrected gates also work even better if the errors of the uncorrected gate are small. In the context of semiconductor spin qubits, for example, corrected gates could in the future be used to correct residual errors from both hyperfine and electrical noise while respecting bandwidth constraints on control pulses. Even in isotopically purified silicon qubits, electrical fluctuations remain a critical source of errors~\cite{yoneda2018quantum}.

\section{Experimental setup}

\begin{figure}
    \centering
    \includegraphics[width=1\linewidth]{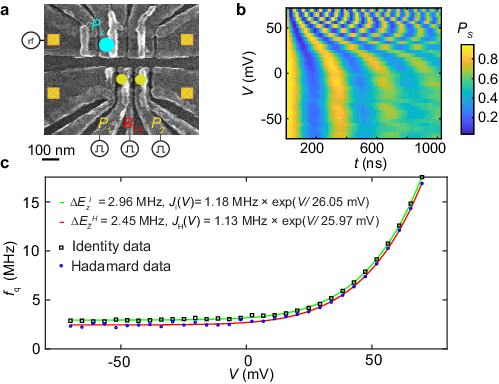}
    \caption{Experimental setup. 
        (a) Silicon double quantum dot similar to the one used in the experiment. The plunger gates $P_1$ and $P_2$ define the qubit (small yellow circles). The voltage $V$ applied to the interdot barrier gate $B_{12}$ controls the exchange coupling between the electrons. During the barrier-gate pulses, compensating pulses are applied to the plunger gates $P_1$ and $P_2$ to keep the dot chemical potential fixed. The plunger gate $P_s$ defines the sensor dot, which is configured for radiofrequency reflectometry. (b) Calibration data used to determine $J$ and $\dez$ for the identity gate. The calibration data for the Hadamard are similar. (c) Voltage-dependent qubit frequencies and corresponding fits to extract $J$ and $\dez$ for the identity and Hadamard experiments. The Identity data are the frequencies extracted from (b).
        }  
    \label{fig:setup}   
\end{figure}

We investigate DCGs in a silicon singlet-triplet (ST) qubit~\cite{burkard2023semiconductor}. Unlike single-spin qubits, which require only time-dependent real or effective magnetic fields for universal control, universal control in ST qubits involves both magnetic and electric fields, making them an ideal platform in which to explore the effectiveness of DCGs in mitigating different forms of noise. Specifically, we use a four-electron ST qubit~\cite{connors2022charge} realized in a double quantum dot fabricated on an undoped, natural-abundance Si/SiGe heterostructure with an overlapping-gate architecture as shown in Fig.~\ref{fig:setup}a. The device is cooled in a dilution refrigerator to a base temperature of approximately 10 mK. The two dots are formed under the plunger gates $P_1$ and $P_2$, and we use the dot under $P_s$ for charge sensing via radiofrequency reflectometry~\cite{connors2020rapid}.  We initialize and measure the qubit via standard Pauli spin-blockade techniques.  

The Hamiltonian for the ST qubit is $H=\frac{h}{2} \left( J(V)\sigma^z+\dez \sigma^x \right)$, where $J(V)$ is the exchange coupling, which depends on the voltage pulse $V$ applied to the barrier gate $B_{12}$.  $\dez$ is the difference in Zeeman energy between the two dots. Most of the experiments discussed below involve pulsing the exchange coupling between the dots. In our device, we expect that $\dez$ results primarily from a $g$-factor difference between the two dots and does not depend significantly on gate voltages~\cite{liu2021magnetic,cai2023coherent}. The two terms in the ST-qubit Hamiltonian are sensitive to different types of noise. Fluctuations in the exchange coupling $\delta J$ result from electrical noise, while fluctuations in the Zeeman gradient $\delta \Delta E_Z$ primarily result from nuclear hyperfine noise.  In the remainder of this work, we concentrate on reducing the infidelity resulting from hyperfine fluctuations. We model hyperfine fluctuations as Gaussian quasistatic noise with a standard deviation $\sigma$. Typical values of $\sigma/\dez$ in our experiment are 0.1-0.2. Values of $\dez$ in our experiments are 2-3 MHz (Fig.~\ref{fig:setup}), corresponding to inhomogeneous dephasing times for singlet-triplet oscillations of $1/(\sqrt{2} \pi \sigma) \approx 800$ ns. These results are consistent with previous reports in natural-silicon devices~\cite{kawakami2014electrical,maune2012coherent}. Values of $\sigma/\dez$ in isotopically purified silicon would be several orders of magnitude lower. 

Before implementing the corrected gates, we calibrate both $\dez$ and $J(V)$. To do this, we perform a Ramsey experiment consisting of a free evolution period at different values of the barrier gate voltage $V$ between $X_{\pi/2}$ pulses generated as $\dez$ rotations.  Figure~\ref{fig:setup}b shows the data from the calibration experiment for the identity gate. To extract both $\dez$ and $J(V)$ from these data, we fit each time series to a decaying sinusoid to extract the voltage-dependent qubit frequency $f_q(V)$. Then, we fit $f_q(V)$ to a phenomenological function $f_q(V) = \sqrt{J(V)^2+\dez^2}$, where $J(V)=J_0 \exp(V/V_0)$, with $J_0$, $V_0$, and $\dez$ as fit parameters. The form of $f_q(V)$ reflects the fact that $J(v)$ and $\dez$ are orthogonal terms in the qubit Hamiltonian, and the form of $J(V)$ reflects the commonly observed exponential dependence of exchange couplings on gate voltages~\cite{martins2016noise,qiao2020coherent}. Figure~\ref{fig:setup}c shows the fits used in this work for the identity gate. We used a similar data set for the Hadamard gate. The values of $\dez$ differ slightly for the two gates because a slightly different magnetic field (between 0.3 and 0.4 T) and device tuning were used. In our device, $g$-factor differences between the two dots determine the value of $\dez$.

\section{Identity gate}

Our goal is to design a DCG that is robust to $\delta\Delta E_Z$ to leading order. In SCQC, the space curve $\vec{r}(t)$ is defined by the first-order error in the evolution operator in the frame defined by the error-free evolution operator $U_0(t)$~\cite{barnes2022dynamically}:
\begin{equation}
\begin{aligned}
    \delta U(t)&= \frac{-ih\delta\Delta E_Z}{2}\int_0^t dt' U_0^\dagger(t')\sigma^x U_0(t')\\&\equiv \frac{-ih\delta\Delta E_Z}{2}\vec{r}(t)\cdot\vec{\sigma}.
\end{aligned}
\end{equation}
Here, the vector $\vec{r}(t)$ of coefficients appearing in this expression can be interpreted as a space curve living in three Euclidean dimensions.
Our goal is to construct an exchange pulse $J(t)$ such that $\vec{r}(t_f)=0=\vec{r}(0)$, i.e., the space curve forms a closed loop, where $t_f$ is the duration of the pulse. This ensures the error cancels out at $t=t_f$. This is in principle possible since $J(t)$ determines $U_0(t)$, which in turn determines $\vec{r}(t)$. The power of SCQC is that it provides a direct connection between $J(t)$ and $\vec{r}(t)$; $hJ(t)$ is proportional to the curvature of the space curve $\vec{r}(t)$. 

This means we can construct a DCG by first constructing a closed space curve and then computing its curvature to obtain the exchange pulse that implements it. We also need to make sure that we achieve the desired target operation, $U_0(t_f)$, which can be guaranteed by imposing suitable boundary conditions on the derivative of the space curve, since 
\begin{equation}\label{eq:boundary_condition}
    \dot{\vec{r}}(t_f)=\frac{1}{2}\hbox{Tr}[U_0^\dagger(t_f)\sigma^x U_0(t_f)\vec{\sigma}].
\end{equation}

\begin{figure}
    \centering
\includegraphics[width=\columnwidth]{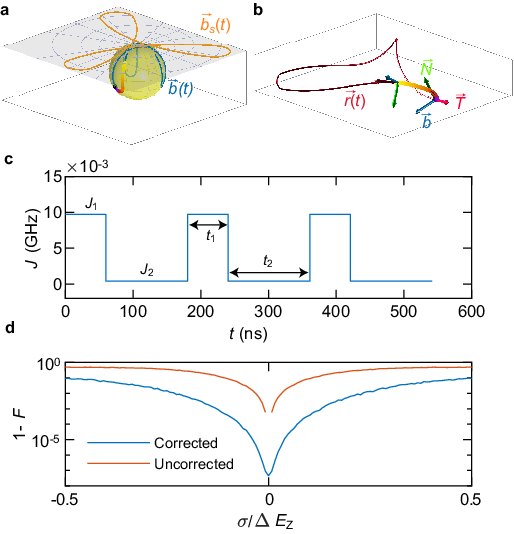}
    \caption{Constructing a dynamically corrected identity gate. (a) To design a closed space curve of constant torsion and positive curvature, we first construct a planar binormal curve $\vec{b}_s(t)$ with three-fold rotational symmetry and then project it onto the unit sphere to obtain a normalized binormal curve $\vec{b}(t)$. The vanishing projected areas of $\vec{b}(t)$ guarantee that the space curve $\vec{r}(t)$ is closed, as shown in (b), where the orthonormal frame defined by the tangent vector $\vec{T}(t)=\dot{\vec{r}}(t)$, the normal vector $\vec{N}=\frac{1}{|\ddot{\vec{r}}|}\dot{\vec{T}}$, and the binormal vector $\vec{b}(t)=\vec{T}(t)\times\vec{N}(t)$ is also shown. (c) The resulting noise-robust exchange pulse is proportional to the geodesic curvature $\kappa_g(t)$ of $\vec{b}(t)$. (d) The predicted infidelity of the corrected and uncorrected gates as a function of noise strength $\sigma$. The uncorrected gate is a 2$\pi$ rotation when $\Delta E_Z \gg J$.
    \label{fig:theory}}
\end{figure}

In addition to its curvature, a space curve in three dimensions is also characterized by a second function called torsion. It was shown in Ref.~\cite{zeng2019geometric} that for the Hamiltonian $H$ above, the torsion is given by $-h\Delta E_Z$. A generic space curve will have non-constant torsion, which means that when constructing $\vec{r}(t)$, we must take care to choose a curve that has constant torsion, in addition to being closed. Another challenge is that we must also make sure that the curvature of $\vec{r}(t)$ remains positive at all times since the exchange coupling is always positive in our device (see Fig.~\ref{fig:setup}c). We can satisfy these constraints by utilizing the systematic procedure for designing closed curves of constant torsion for SCQC presented in Ref.~\cite{zhuang2022noise}. The idea is to express the space curve in terms of its \emph{binormal} curve $\vec{b}(t)$:
\begin{equation}\label{eq:r_from_B}
    \vec{r}(t)=-\frac{1}{h\Delta E_Z}\int_0^tdt'\vec{b}(t')\times\dot{\vec{b}}(t'),
\end{equation}
where $\vec{b}(t)$ is a unit vector: $|\vec{b}(t)|=1$. Any $\vec{b}(t)$ will generate a space curve of constant torsion using the above formula. Each component of the integral in Eq.~\eqref{eq:r_from_B} is in fact the area enclosed by the curve after projecting it onto a Cartesian plane orthogonal to that component. This means that we can ensure that the space curve closes by constructing a $\vec{b}(t)$ that has vanishing-area projections. Following Ref.~\cite{zhuang2022noise}, we can do this by first constructing a planar curve $\vec{b}_s(t)=2\left(\frac{b_x(t)}{1+b_z(t)},\frac{b_y(t)}{1+b_z(t)}\right)$ with a discrete rotational symmetry and then stereographically projecting (from the south pole) the result onto the surface of a unit sphere: $\vec{b}(t)=(b_x(t),b_y(t),b_z(t))$. The rotational symmetry guarantees that the projected areas of the spherical curve vanish. This procedure for constructing $\vec{b}(t)$ is illustrated in Fig.~\ref{fig:theory}a, while the resulting closed space curve $\vec{r}(t)$ is shown in Fig.~\ref{fig:theory}b. In this example, the boundary conditions $\vec{b}(t_f) = \vec{b}(0)$ and $\dot{\vec{b}}(t_f) = \dot{\vec{b}}(0)$, which ensure the curve closes smoothly ($\dot{\vec{r}}(t_f)=\dot{\vec{r}}(0)$), yield an identity gate. We can also extract the corresponding exchange pulse directly from $\vec{b}(t)$ by computing its geodesic curvature:
\begin{equation}\label{eq:geodesic_curvature}
\kappa_g(t)=\frac{\ddot{\vec{b}}\cdot(\vec{b}\times\dot{\vec{b}})}{|\dot{\vec{b}}|^3}=\frac{J(t)}{|\Delta E_Z|}.
\end{equation}
The geodesic curvature gives the exchange coupling as a function of time for the error-corrected identity gate as shown in Fig.~\ref{fig:theory}c. Based on the underlying three-fold symmetry of $\vec{b}_s(t)$, the gate alternates between two values of $J$, which we refer to as $J_1$ and $J_2$, for different amounts of time $t_1$ and $t_2$, respectively. The values of these parameters depend on $\dez$. For the experiments on the identity gate described below, $\dez=$ 2.9 MHz, so $J_1=$ 9.7 MHz, $J_2=$ 0.4 MHz, $t_1=60$~ns, and $t_2=121$~ns. The expected infidelity of the gate is shown in Fig.~\ref{fig:theory}d. 

A major benefit of our gate construction, compared with other types of corrected gates, is the relatively small number of parameters ($J_1$, $J_2$, $t_1$, and $t_2$) required to describe the pulses. As a result, we can sweep the parameters to calibrate the gate in a reasonable amount of time. Empirically, we find that sweeping $J_1$ and $J_2$ alone can relatively quickly yield gates with high fidelity, as discussed further below. Including the coarse tuning discussed in the previous section and the fine-tuning discussed here, the entire tune-up process takes a few hours. We usually implement the full tune-up process each day.   % To initially calibrate $\dez$ and $J(V)$, we perform a Ramsey experiment consisting of a free evolution period at different values of the gate voltage $V$ between $X_{\pi/2}$ pulses generated as $\dez$ rotations.  Figure~\ref{fig:setup}b shows the data from this calibration experiment. To extract both $\dez$ and $J(V)$ from these data, we fit each time series to a decaying sinusoid to extract the voltage-dependent qubit frequency $f_q(V)$. Then, we fit $f_q(V)$ to a phenomenological function $f_q(V) = \sqrt{J(V)^2+\dez^2}$, where $J(V)=J_0 \exp(V/V_0)$, with $J_0$ and $V_0$ as fit parameters.  Figure~\ref{fig:setup}c shows the fits used in this work.

To assess the performance of the gate, we perform self-consistent standard quantum process tomography. To account for state-preparation and measurement errors, as well as tomographic rotation errors, we follow the calibration procedures described in Refs.~\cite{shulman2012demonstration,takahashi2013tomography,nichol2017high}. In brief, we use a fitting process that self-consistently finds the positive operator-valued measure (POVM) operators that describe our tomographic measurements. Our primary assumption in this fitting process is that the purity of the qubit state should smoothly decay in time as a result of dephasing. Tomographic errors, for example, can result in a state purity that appears to oscillate rapidly in time if the POVM operators are imperfectly calibratrated~\cite {shulman2012demonstration}. Once the POVM operators are calibrated, we measure both the input and output states for the specific DCG configuration, using a maximum likelihood method to determine the most likely physical density matrices~\cite{smolin2012efficient}. Through a constrained direct inversion process, we find the most-likely physical process matrix corresponding to the gate and the resulting gate fidelity $F.$ See Appendix A for further information on our tomographic calibration procedure. 

\begin{figure}
    \centering
    \includegraphics[width=\linewidth]{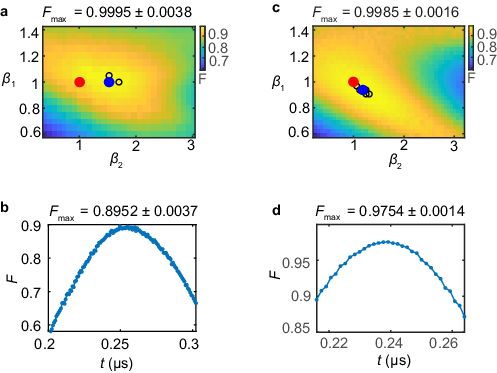}
    \caption{Process fidelity result for identity gate. 
    (a) Measured fidelity for the corrected gate. $\beta_1$ and $\beta_2$ are scaling factors for the two exchange couplings which parameterize the gate. The red dot indicates the predicted optimal configuration $\beta_1=\beta_2=1$, and the blue dot corresponds to the observed configuration with the maximum fidelity. The black symbols indicate other configurations with a fidelity equal to the maximum within the uncertainty.
    (b) Measured fidelity of an uncorrected identity gate. $t$ is the gate duration. 
    (c) Simulated fidelity of the corrected identity gate. %$t$ is the gate duration. 
    (d) Simulated fidelity of the uncorrected identity gate. }
    \label{fig:identity}
\end{figure}

To characterize the corrected identity gate, we sweep $\beta_1=J_1'/ J_1$ and $\beta_2=J_2'/ J_2$, where $J_1'$ and $J_2'$ are the actual exchange values during the gate, and $J_1$ and $J_2$ are the theoretically predicted values. We measure the process fidelity at each value of $\beta_1$ and $\beta_2$ (Fig.~\ref{fig:identity}a). The observed optimal configuration is similar to the predicted configuration, and it has an optimal fidelity larger than $0.99$, as shown in Fig.~\ref{fig:identity}. The measured performance of the gate also agrees with numerical simulations that include the finite timing resolution of our arbitrary waveform generator, quasistatic hyperfine noise, quasistatic charge noise, and pulse distortions, which arise from imperfections in our setup (discussed further below), as shown in Fig.~\ref{fig:identity}c. The simulations assume perfect state preparation and readout. A main cause for the difference between the predicted optimal configuration ($\beta_1=\beta_2=1$) and the observed optimal configuration involves pulse distortions in our setup, which we discuss in detail below. We also measure the fidelity of an uncorrected identity gate (Fig.~\ref{fig:identity}b), implemented as a 2$\pi$ rotation when $J=\dez$, and we find that the corrected identity gate reduces the infidelity by more than a factor of 10. The corresponding simulation for the uncorrected identity gate is shown in Fig~\ref{fig:identity}d.

\section{Hadamard gate}
\begin{figure}
    \centering
\includegraphics[width=\columnwidth]{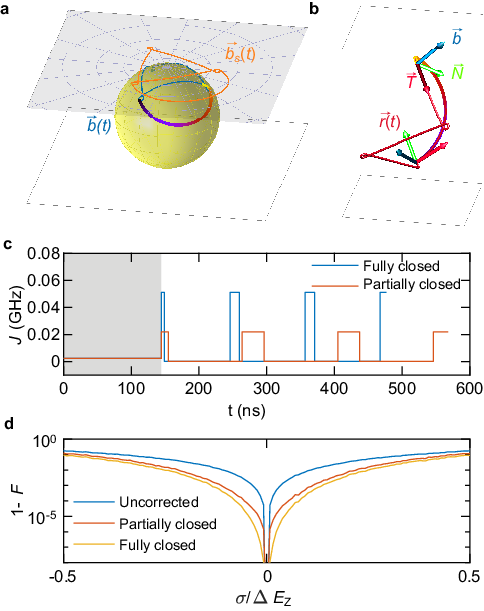}
    \caption{Constructing a dynamically corrected Hadamard gate. (a) A planar binormal curve $\vec{b}_s(t)$ comprised of a piece with three-fold rotational symmetry that generates a noisy identity gate connected to an arc that generates a noisy Hadamard. The stereographic projection of $\vec{b}_s(t)$ onto the unit sphere, $\vec{b}(t)$, is also shown. (b) The corresponding closed space curve $\vec{r}(t)$. (c) The resulting noise-robust exchange pulse is proportional to the geodesic curvature $\kappa_g(t)$ of $\vec{b}(t)$. The shaded area corresponds to the noisy Hadamard, while the unshaded area generates an identity gate whose error cancels that of the Hadamard gate. (d) Expected infidelity of the gate as a function of hyperfine noise strength.}
    \label{fig:hadamard_theory}
\end{figure}

To design a non-identity gate, we need to construct a closed space curve of constant torsion that has a cusp at the origin. For example, in the case where $U_0(t_f)$ is chosen to be a Hadamard gate, we have $\dot{\vec{r}}(0)=\hat x$, while from Eq.~\eqref{eq:boundary_condition} we have $\dot{\vec{r}}(t_f)=\hat z$. The boundary condition for the space curve is $(\vec{T},\vec{b})_{t=t_f} = (\vec{b},\vec{T})_{t=0}$. 
We again use the method of Ref.~\cite{zhuang2022noise} to construct a binormal curve and impose the boundary conditions on $\vec{b}$ and its first derivative at $t=0$ and $t=t_f$ to ensure a Hadamard gate is generated. To construct a suitable curve, we follow the approach of Refs.~\cite{brown2004arbitrarily,wang2012composite,kestner2013noise} in which errors are cancelled by concatenating a noisy target gate with a noisy identity gate with equal but opposite error. In this spirit, we design $\vec{b}(t)$ to consist of segments corresponding to a noisy Hadamard gate and segments corresponding to a noisy identity gate. Our explicit construction is shown in Fig.~\ref{fig:hadamard_theory} along with the resulting exchange pulse.

While in principle this approach should enable high-fidelity gates, we find that implementing these gates can require exchange couplings outside the range of possible values for this device (Fig.~\ref{fig:hadamard_theory}c). One way to solve this challenge is to generate gates with longer times and smaller exchange values. Another approach is to relax the condition for complete closure of the space curve. Empirically, we find that this process can generate gates with smaller exchange pulses and relatively short times. The resulting partially-closed gate is shown in Fig.~\ref{fig:hadamard_theory}c, and the expected infidelities are shown in Fig.~\ref{fig:hadamard_theory}d. While the partially-closed gate does not perform quite as well as the fully-closed gate, it requires exchange pulses only about half as large as the fully-closed gate.  The resulting gate is parameterized by three exchange values: $J_H$, $J_1$, and $J_2$, and four time values $t_H$, $t_1$, $t_2$, and $t_b$, and the actual gate sequence consists of exchange pulses $(J_H, J_1, J_2, J_1, J_2, J_1, J_2,J_1)$ for times $(t_H, t_1-t_b, t_2, t_1, t_2, t_1,t_2, t_b)$. For these experiments, $\dez=2.5$ MHz. $J_H=\dez$ and $t_H=1/(2\sqrt{2}\dez)$ are the noisy Hadamard gate parameters, which we optimize through a calibration routine (see Appendix A). In addition, $J_1=22$ MHz, $J_2=$ 0.1 MHz, $t_b=21.7$ ns, $t_1=32$ ns, and $t_2=109$ ns.

\begin{figure}
    \centering
    \includegraphics[width=\linewidth]{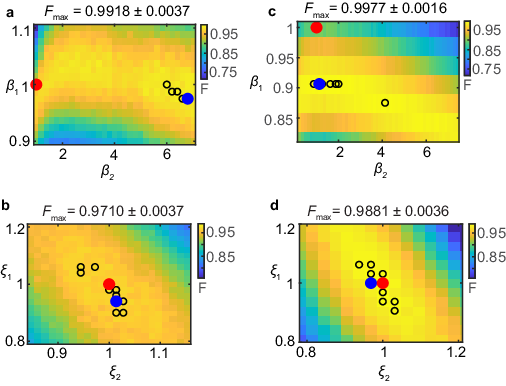}
    \caption{Process fidelity results for Hadamard gate. 
        (a) Measured fidelity for the corrected gate. The red dot indicates $\beta_1=\beta_2=1$, and the blue dot corresponds to the observed configuration with the maximum fidelity. The black symbols indicate other configurations with a fidelity equal to the maximum within the uncertainty.  
        (b) Measured fidelity for the uncorrected Hadamard gate. For each configuration, $J=\xi_1\dez$ and the gate duration $t=\xi_2 /(2\sqrt{2}\dez)$.
        (c) Simulated fidelity of the corrected Hadamard gate. 
        (d) Simulated fidelity of the uncorrected Hadamard gate.}
    \label{fig:hadamard}
\end{figure}

To optimize the corrected Hadamard gate, we sweep $\beta_1$ and $\beta_2$ as before, and we measure the process fidelity at each value of $\beta_1$ and $\beta_2$ (Fig.~\ref{fig:hadamard}a).  For a range of parameter configurations that differ slightly from the theoretically predicted values, we find gate fidelities $>0.99$. Experimentally we find a maximum process fidelity of $0.9918\pm0.0037$. To assess how well the DCG corrects for errors, we measure the process fidelity of the uncorrected Hadamard gate, implemented as a square pulse with $J=\dez$. In this case, since the magnetic field determines the mean value of $\dez$, we sweep both $\xi_1=J/J_H$ and the gate duration $\xi_2=t/t_H$, where $J$ is the actual value of the exchange coupling during the gate, and $t$ is the gate time. We find a maximum uncorrected fidelity of about $0.9710\pm0.0037$ (Fig.~\ref{fig:hadamard}b). Thus, the DCG reduces the infidelity by about a factor of three. We compute uncertainties in the fidelities by fitting the measured fidelities versus the scaling factors to smooth polynomials. Then we take the standard deviations of the distributions of the residuals. For numerical simulations (discussed below), we find that this method appears not to underestimate the actual standard deviations of the distributions of fidelities over multiple runs with the same configuration. See the Appendix B for further details on the uncertainty estimation.  

\begin{figure}
    \centering
    \includegraphics[width=\linewidth]{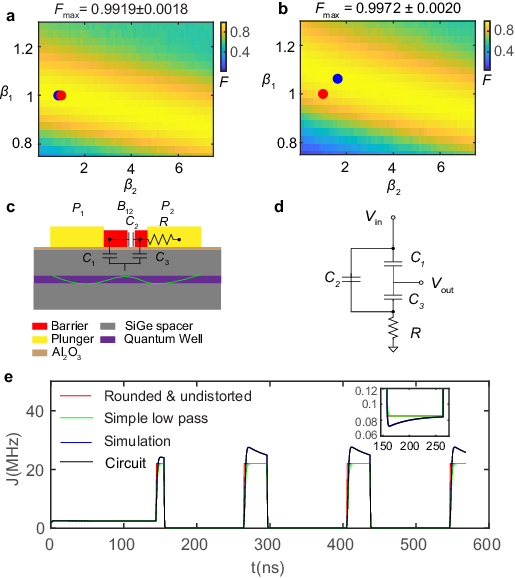}
    \caption{Effects of pulse distortions. 
        (a) Simulated Hadamard gate fidelity without pulse distortion. 
        (b) Simulated Hadamard gate fidelity with low-pass pulse distortion. 
        (c) Cartoon of a possible broken gate and its circuit representation. While the drawing is not to scale, the typical lateral width of a gate is about 50 nm.
        (d) Circuit model that reproduces the pulse distortions used in our simulations. The component values that match the hypothesized distortion are $R=10$ G$\Omega$, $C_1=1$ aF, $C_2=4$ aF, and $C_3=0.05$ aF.  
        (e) Pulse sequences used in simulating (a) (red), (b) (green), and the output of our circuit model (d) (black). Overlapping the black curve is a blue curve that implements the partial high pass filter in our simulations (see Appendix C). The black curve has been rescaled to account for the capacitances in the circuit. Such a rescaling would be accounted for in our calibration of $J(V)$.  }
    \label{fig:pulse}
\end{figure}

We also numerically simulate the performance of the Hadamard gate. Unsurprisingly, simulations without any pulse distortions (Fig.~\ref{fig:pulse}a) disagree qualitatively with our data. In particular, the measured high-fidelity region bends down at small values of $\beta_2$ (Fig.~\ref{fig:hadamard}a), but the simulation (Fig.~\ref{fig:pulse}a) does not show this effect. At first, one might suspect the presence of an effective low-pass filter in our setup, which would result from stray capacitances and resistances in our cryostat wiring. However, simulations including only an effective low-pass filter with a time constant of a few nanoseconds also fail to qualitatively match our data (Fig.~\ref{fig:pulse}b). Such a model gives a result that slightly curves up at small values of $\beta_2$. One pulse distortion model that yields a downward curve consists of a ``partial'' high-pass filter (Fig.~\ref{fig:pulse}c,d) (see the Appendix C). We envision that such a scenario could result from a partially broken gate electrode.  While we cannot definitively say that one of our gates is broken, this scenario is plausible because the grain size of the Al film used to create the gates approaches the size of the gates themselves. Moreover, this model and the circuit parameters we use to generate this pulse distortion yield simulation results consistent with our data (Figs.~\ref{fig:hadamard}c,d) and physical expectations for a broken gate. Figure~\ref{fig:pulse}e shows both the undistorted Hadamard pulse sequence as well as the Hadamard sequence distorted by the circuit discussed above. We expect that the differences between the simulated and observed gate fidelities shown in Figs.~\ref{fig:identity} and \ref{fig:hadamard} have to do with properties of the pulse distortions not captured by our model.  

\section{Discussion}
One notable feature of our results is a larger improvement in gate fidelity for the identity than the Hadamard. In part, this occurs because the measured uncorrected identity gate has a lower fidelity than the simulated uncorrected gate. The reason for this discrepancy is unclear and is the subject of ongoing work. It may have to do with possible correlated effects of hyperfine fluctuations on both state preparation and readout and the gate itself (see Appendix A) or a potential violation of the quasistatic noise assumption during experimental conditions. In addition, the fidelity of the corrected identity gate is also considerably higher than the fidelity of the corrected Hadamard gate. The imperfect cancellation of errors in our Hadamard construction likely contributes to this effect. 

\begin{table}[h]
\centering
\begin{tabular}{c|cccc}
\hline
\hline
Gate & $\dez \& J$ & $\dez$  & $J$ & No noise \\ \hline
H& 0.9881&0.9888&0.9996&1\\

DCG H&0.9977&0.9987&0.9981&0.9999\\

 I&0.9754&0.9760&0.9994&0.9999\\

DCG I &0.9985&0.9989&0.9994&1\\
Undistorted DCG H & 0.9974&0.9989&0.9983&1\\
Undistorted DCG I &0.9992&0.9996 & 1&1\\
\hline
\hline
\end{tabular}
\caption{Simulation of the effects of charge noise and hyperfine noise on DCG performance. Each number is a simulated fidelity for the gate indicated by the row label with the noise configuration indicated with the column label.  In the labels, ``H'' and ``I'' refer to the uncorrected Hadamard and identity gates, respectively. The ``undistorted'' gates refer to gates without pulse distortions and with infinitely precise timing resolution. In the column labels, $\dez$ represents hyperfine noise and $J$ represents charge noise. In the column labeled $\dez \& J$, for example, both noise sources are taken into account. }
\label{table:1}
\end{table}

To assess the contribution of the different noise sources to the overall infidelity of the DCGs, we conducted simulations with different combinations of charge noise and hyperfine noise, as shown in Table~\ref{table:1}. For the uncorrected gates, dephasing due to hyperfine noise contributes the largest error. For the corrected gates, dephasing due to both hyperfine and electrical noise appears to contribute roughly equally to the infidelity. Based on these simulations, it also appears that the fidelity of the corrected Hadamard gate is not strongly affected by pulse distortions in our setup. On the other hand, it appears that the identity gate could perform even better without pulse distortions. The reason for this difference in sensitivity to pulse distortions is not clear and will be the subject of future work. The last column under ``No noise'' confirms perfect fidelity in the absence of noise.

In summary, we have described the construction and implementation of corrected Hadamard and identity gates in a silicon singlet-triplet qubit. Because of the small number of parameters describing the pulses, we optimized the gates by sweeping the control parameters, finding improved fidelities for both the Hadamard and identity gates. The results demonstrate the effectiveness of the SCQC approach for designing the gates, as well as the importance of pulse distortions on the overall gate fidelity. For the gates we have explored here, areas of future work include investigating the difference in sensitivity to pulse distortions between the Hadamard and identity gates and varying other pulse parameters, such as the timing, to optimize the gates further.

We also envision that our results will motivate future work on implementing corrected gates to correct for other types of errors, such as those originating from electrical noise in semiconductor spin qubits~\cite{nelson2023designing}. Corrected gates of the type we have explored can also be implemented as smooth pulses to respect control-pulse bandwidth restrictions~\cite{zeng2018general} and even correct for pulse errors~\cite{nelson2023designing}. We also emphasize that the relative improvement in gate fidelity generally increases with the fidelity of the uncorrected gate (Figs.~\ref{fig:theory}d and \ref{fig:hadamard_theory}d). Thus, we expect that the gates that we describe here should perform even better in isotopically purified spin qubits. As a result, corrected gates will likely complement other techniques in the effort to lower gate errors in quantum computing devices. 

\section*{Acknowledgments}

EB acknowledges support from the Office of Naval Research (Grant No. N00014-21-1-2629). JMN, HYW, and XXC acknowledge support from the Army Research Office (Grant Nos. W911NF-17-1-0260 and W911NF-19-1-0167), and the National Science Foundation (Grant Nos. DMR-1941673 and OMA 1936250). EB, JMN, and HYW acknowledge support from the Department of Energy (Grant No. DE-SC0022389).
\section{Appendix A: State and Process Tomography}
To perform state tomography for experiments on the identity gate, we used the following operations to project the qubit state along the $x$, $y$, and $z$ axes of the Bloch sphere:  $H$, $H H' H^{1/2}$, and $H^2$. Here, $H$ represents a Hadamard gate, and $H'$ represents a $\pi$ rotation when the direction of the effective field represented by the Hamiltonian points at 22.5$^{\circ}$ away from the $z$ axis, compared with the usual Hadamard, which points at $45^{\circ}$. Both gates were calibrated with a process similar to that described in Ref.~\cite{connors2022charge}. While the corrected identity gate is minimally sensitive to hyperfine fluctuations, the state preparation operations, which rely heavily on the Hadamard gate, are sensitive to hyperfine noise. We therefore suspect that the uncorrected identity gate's relatively low fidelity could result from hyperfine errors that affect both the state preparation and gate. Our simulations do not include this effect and assume perfect state preparation and readout. 

\begin{figure}
    \centering
    \includegraphics[width=0.5\linewidth]{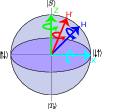}
    \caption{Operations used to calibrate gates and perform state and process tomography.}
     \label{fig:BlochSphere}
\end{figure}

\begin{figure*}
    \centering
    \includegraphics[width = 1\linewidth]{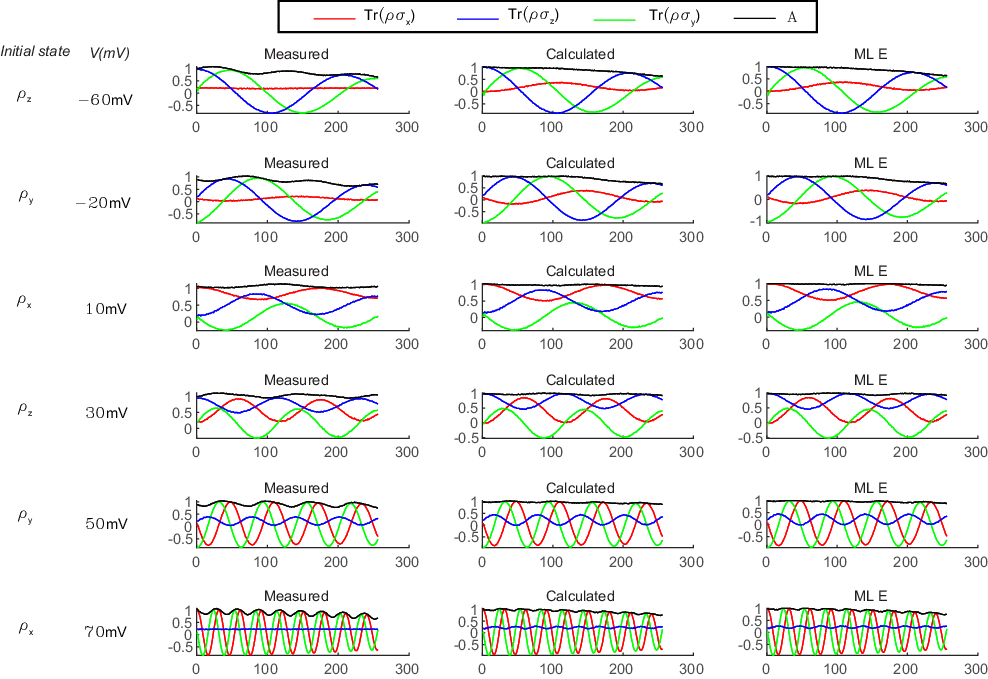}
    \caption{State-tomography calibration of the measurement operators used for the Hadamard gate. We prepare multiple different initial states, and we evolve these initial states under different exchange coupling strengths. By measuring and fitting the data at different times, we can extract the parameters of our tomographic measurements. Each row of the table above corresponds to a different initial state/control-field combination. The ``Initial state'' column lists the approximate initial state on the Bloch sphere (Fig.~\ref{fig:BlochSphere}), and the ``V'' column lists the barrier gate voltage applied during the evolution. At different evolution times, we perform tomographic measurements of the states.  The column marked ``Measured'' shows the tomographic reconstruction of the state without calibration. The state purity $A$ oscillates in time, violating our assumption that it should smoothly decay because of dephasing and relaxation. To correct for this, we perform a least-squares fit of the POVM operators to minimize the oscillations in the state purity. The column marked ``Calculated'' shows the tomographic reconstructions with the fitted POVM operators, and the state purity decays more smoothly than the ``Measured'' case. The ``MLE'' column represents the most-likely physical density matrix~\cite{smolin2012efficient} corresponding to the ``Calculated'' density matrices.}
    \label{fig:TomocalH}
\end{figure*}

To perform state tomography for experiments on the corrected Hadamard gate, we used an adiabatic readout to measure along the $x$ direction, a $Z_{\pi/2}$ gate followed by an adiabatic readout to measure along the $y$ direction, and an adiabatic readout to measure along the $z$ direction. We avoided using Hadamard gates during the state tomography operations to minimize correlated errors during process tomography. We calibrated our state tomography by fitting the values of the POVM operators describing our tomographic measurements~\cite{takahashi2013tomography} using a process similar to that described in Ref.~\cite{shulman2012demonstration}. Figure \ref{fig:BlochSphere} shows a Bloch-sphere schematic of the operations used in the state preparations. The tomographic calibration used for the Hadamard gate is shown in Fig~\ref{fig:TomocalH}. Similar calibration data are used for the identity gate.

The initial states for process tomography were prepared with operations similar to our readout operations with one additional initial state that is needed for normalization. The qubit states along $-y$ and $-x$ axes of the Bloch sphere were used for the identity and Hadamard gate respectively. These respectively were prepared using $H^{3/2}H'H$ and an adiabatic ramp followed by a $Z$ gate. We calculated process matrices using a constrained maximum likelihood approach using a process similar to that described in Ref.~\cite{nichol2017high}.
\section{Appendix B: Error bars}
\begin{figure*}
    \centering
    \includegraphics[width=1\linewidth]{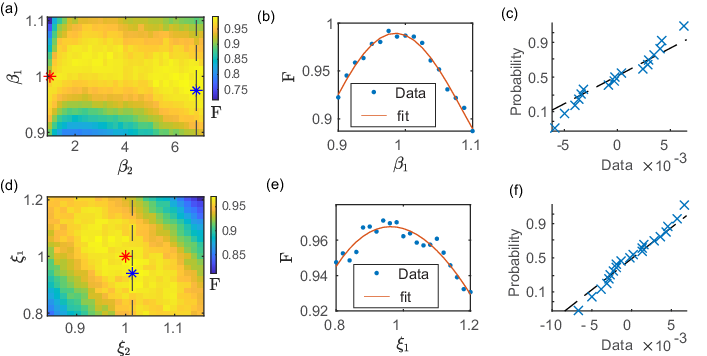}
    \caption{Extraction of the uncertainties for the Hadamard gate. (a) Data for the corrected Hadamard and linecut. (b) Linecut data and fit to a quartic function. (c) Cumulative probability plot for the data in (b), showing that the distribution of the residuals (blue crosses) can be approximated by a normal distribution (dashed line). %to test for normality of the residuals of the data in b. 
    (d) Data for the uncorrected Hadamard and linecut. (e) Linecut data and fit to a quartic function. (f) Cumulative probability plot for the residuals in e. }
    \label{ErrorbarH}
\end{figure*}

\begin{figure*}
    \centering
    \includegraphics[width=1\linewidth]{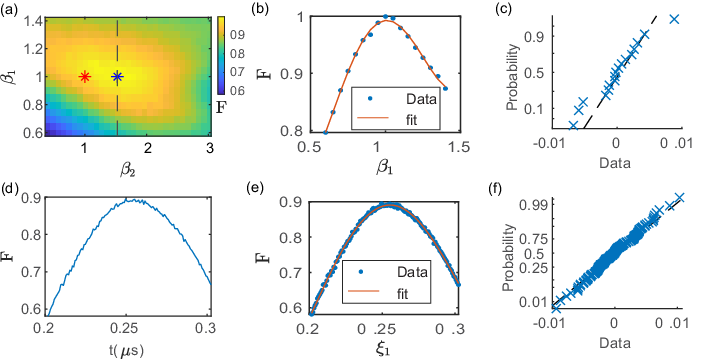}
    \caption{Extraction of the uncertainties for the identity gate. (a) Data for the corrected identity and linecut. (b) Linecut data and fit to a quartic function. (c) Probability plot to test for normality of the residuals of the data in b. (d) Data for the uncorrected identity and linecut. (e) Linecut data and fit to a quartic function. (f) Probability plot for the residuals in e. }
    \label{ErrorbarI}
\end{figure*}

We estimate uncertainties for our extracted gate fidelities by fitting line cuts of our measured fidelities to smooth polynomials. We assume that deviations from the smooth curve represent Gaussian random errors in our procedure, and we report the uncertainty as the standard deviation of the distribution of residuals.Shown in Figs.~\ref{ErrorbarH} and \ref{ErrorbarI} (a)-(e) are the linecuts and fits for the corrected Hadamard and identity gates from the experiment, respectively. The probability plot, which compares the cumulative distributions of the residuals with a normal distribution, in Figs.~\ref{ErrorbarH} (c) and (f) is to show that the residuals can be approximated as normally distributed and hence the standard deviation can be used as the error bar.

\begin{figure*}
    \centering
    \includegraphics[width=1\linewidth]{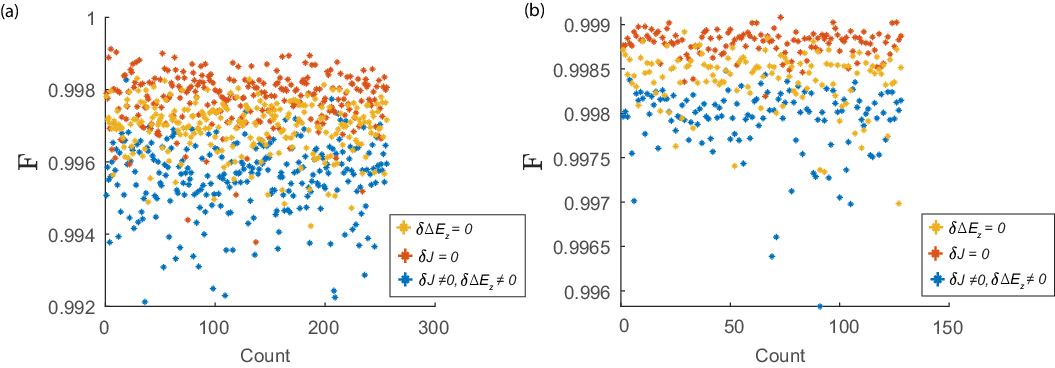}
    \caption{Scatter plot of the fidelity for different noise realizations of the optimal gate, from simulations of (a) DCG Hadamard and (b) DCG identity gate. The different colors are for different noise conditions. Red corresponds to hyperfine noise only, orange corresponds to charge noise only, and blue is when both noises are considered. Respectively for red, orange, and blue colors, the fidelities are $0.9977 \pm 0.0007$, $0.9969  \pm 0.0007$, $0.9955 \pm 0.001$ for the Hadamard gate and $0.9988 \pm 0.0001$,$0.9984 \pm 0.0003$, $0.9979 \pm 0.0004$ for identity gate. The Hadamard gate is averaged over 128 noise realizations while the identity gate is averaged over 256 realizations to match the experimental conditions.}
    \label{fig:Dist}
\end{figure*}

To justify this approach for estimating the uncertainties, we simulated the measurements and confirmed that the standard deviation extracted in the way we have discussed agrees reasonably with the standard deviation of the distribution of fidelities of a single configuration. Figure~\ref{fig:Dist} shows the simulated distribution of fidelities of optimal Hadamard (a) and identity (b) gates. For the Hadamard gate, the fitted uncertainty (0.0016, see Fig. 5c in the main text) is a factor of 1.6 larger than the actual standard deviation (0.001), while for the identity gate, the fitted uncertainty (0.0016, see Fig. 3c in the main text) is about a factor of 4 larger than the actual standard deviation (0.0004). Figure~\ref{fig:Dist} also shows the expected gate performance of the three different noise configurations shown in Table I of the main text.
\section{Appendix C: Simulation}
Our simulations assume quasistatic hyperfine and charge noise, pulse distortions due to our setup, the temporal resolution of our arbitrary waveform generator, and perfect state preparations and measurements. We assumed $\sigma= 0.2867$ MHz, corresponding to a $T_2^*= $785 ns of $\dez$ rotations. We assumed fractional exchange noise, such that $\sigma_J/J = 0.012$, corresponding to an exchange oscillation quality factor of about 18, appropriate for barrier-controlled exchange pulses. Except for the last two rows of Tabe I in the main text, all segment times are rounded to the nearest whole nanosecond. We use a 1~ns time step for our simulations, and we solve the time-independent Schr\"odinger equation for each time step. 

To implement the pulse distortions used in our simulation, we used a low-pass filter with time constant $\tau_{lp}=1$ ns and kernel $K_{lp}(t)=\exp(-t/\tau_{lp})/\tau_{lp}$, and a ``partial'' high pass filter with amplitude $A_{hp}=0.05$, time constant $\tau_{hp}=40$ ns, and kernel $K_{hp}(t)=A_{hp} \left( \delta(t) -\exp(-t/\tau_{hp})/\tau_{hp} \right)$. 
We implemented the pulse distortion for a given gate sequence by first creating a voltage time series $V(t)$ using the exchange-to-voltage conversions discussed in the main text. Then, we created a distorted waveform $V'(t)=V \circ K_{lp} \circ (1 + K_{hp})$, where $\circ$ indicates a convolution, and $1$ indicates an identity operation by convolution with the Dirac delta function. We converted this distorted voltage waveform back to a distorted exchange waveform and solved the Schr\"odinger equation as described above. We envision that the low-pass filter results from stray resistances or capacitances in our setup, and the partial high-pass filter could result from a broken gate. The circuit discussed in the main text yields actual waveforms that are highly similar to the waveforms generated using the filtering process described here. 

In all cases, the reported fidelities are extracted from process matrices calculated using a maximum likelihood estimation process using the CVX optimization package, where the Choi matrices were constrained to be completely positive and trace-preserving.
%\bibliography{bib2}
\end{document}